\documentstyle[aps,prl,preprint,epsfig]{revtex}
\begin{document}
\draft
\title{Incommensurate and commensurate antiferromagnetic spin fluctuations in
$Cr$ and $Cr$-alloys from ab-initio dynamical spin susceptibility calculations}
\author{J.B.Staunton\dag, J.Poulter\ddag, B.Ginatempo\P, E.Bruno\P and
D.D.Johnson\S}
\address{\dag\ Department of Physics, University of Warwick, Coventry CV4~
7AL, U.K.}
\address{\ddag\ Department of Mathematics,Faculty of Science,
Mahidol University,Bangkok 10400,
Thailand.}
\address{\P\ Dipartimento di Fisica and Unita INFM, Universita di Messina, Italy}
\address{\S\ Department of Materials Science and Engineering, University of
Illinois, IL 61801, U.S.A.}
\date{\today}
\maketitle
\begin{abstract}
A scheme for making {\it ab-initio} calculations of the dynamic paramagnetic spin
susceptibilities of solids at finite temperatures is described. It is based on
Time-Dependent Density Functional Theory and employs an electronic multiple
scattering formalism. Incommensurate and commensurate anti-ferromagnetic
spin fluctuations in paramagnetic $Cr$ and compositionally disordered
$Cr_{95}V_{5}$ and $Cr_{95}Re_{5}$ alloys are studied together with the
connection with the nesting of their Fermi surfaces. We find that the spin
fluctuations
can be described rather simply in terms of an overdamped oscillator model.
Good agreement with inelastic neutron scattering data is obtained.
\end{abstract}
Chromium is the archetypal itinerant anti-ferromagnet (AF) whose
famous incommensurate spin density wave (SDW)
ground state is determined by the nesting
wave-vectors ${\bf q}_{nest}$ identified in the Fermi surface \cite{Fawcett}. Chromium alloys also have varied AF properties
\cite{Fawcett} and their
paramagnetic states have recently attracted attention owing, in part, to analogies
drawn with the high temperature superconducting cuprates especially
$(La_{c}Sr_{1-c})_{2}CuO_{4}$.
The incommensurate SDW fluctuations in these materials
\cite{Mason1992} are rather similar to those seen in the paramagnetic phase of $Cr$
close to $T_{N}$. Moreover `parent' materials $La_{2}CuO_{4}$ in the one
instance and $Cr_{95}Mn_{5}$ or $Cr_{95}Re_{5}$ in
the other are simple commensurate AF materials which on
lowering the electron concentration by suitable doping develop incommensurate spin
fluctuations which may be promoted by imperfectly nested Fermi surfaces.

Here we
examine the nature of damped diffusive spin fluctuations in
chromium above the Ne$\acute{e}$l temperature $T_{N}$ which are precursory to the
SDW ground state. We also study dilute
chromium alloys, $Cr_{95}Re_{5}$ and $Cr_{95}V_{5}$ and obtain good agreement with
experimental data. For example, recent inelastic neutron scattering
experiments \cite{Fawcett,Hayden} have measured incommensurate AF
`paramagnons', persisting up to high frequencies in the latter system.
We explore the temperature dependence, variation with dopant concentration
and the evolution of the spin fluctuations in these systems from
incommensurability to
commensurability with increasing frequency and provide the first ab-initio
description of these effects. To this end we describe a new scheme for calculating
the wave-vector and frequency dependent dynamic spin susceptibility of metals which
is based on the Time Dependent Density Functional Theory (TDDFT) of Gross et al.
\cite{Gross_review} and as such is an all electron theory. For the first time the
temperature dependent dynamic spin susceptibility of metals and alloys is
calculated from this basis.
There have been several simple parameterised models
to describe the magnetic properties of $Cr$ and its alloys
\cite{model}. All of these have
concentrated on the approximately nested electron `jack' and slightly larger
octahedral hole pieces of the Fermi surface \cite{Fawcett}
and, at best, have only included the effects
of all the remaining electrons via an electron reservoir. We find similarities
between our results and results from such models but show that a complete picture
is obtained only when an electronic band-filling effect which favors a simple AF
ordering at low temperature is also considered. We also find that the spin
fluctuations are given an accurate description as overdamped diffusive simple
harmonic oscillator modes which are at the heart of theories of the effects of spin
fluctuations upon the properties of itinerant electron systems \cite{GGLetc}.

Over the past few years great progress has been made
in establishing TDDFT \cite{Gross_review}. Analogs of the
Hohenberg-Kohn \cite{HK+KS} theorem of the static density
functional formalism have been proved and rigorous properties found.
Here we consider a
paramagnetic metal subjected to a small, time-dependent
external magnetic field, ${\bf b}({\bf r},t)$ which induces a
magnetisation ${\bf m} ({\bf r},t)$ and
use TDDFT to derive an expression for the dynamic
paramagnetic spin susceptibility $\chi ({\bf q}, w)$ via
a variational linear response approach \cite{Yang}.
Accurate calculations of dynamic susceptibilities from this
basis are scarce (e.g. \cite{SW+Sav}) because
they are difficult and computationally demanding. Here we mitigate these
problems by  accessing $\chi ({\bf q}, w)$ via the
corresponding {\it temperature} susceptibility $\bar{\chi} ({\bf q}, w_{n})$
where $w_{n}$ denotes a bosonic Matsubara frequency \cite{Fetter+Walecka}. We
outline this approach below.

The equilibrium state of a paramagnetic metal, described by standard DFT,
has density $\rho_{0}({\bf r})$ and its
magnetic response function $\chi ({\bf r} t;{\bf r}^{\prime}t^{\prime}) =
(\delta m [b] ({\bf r},t)/\delta b ({\bf
r}^{\prime},t^{\prime}))|_{b=0,\rho_{0}}$ is given by the following Dyson-type
equation.
\begin{equation}
\chi ({\bf r}t;{\bf r}^{\prime}t^{\prime}) = \chi_{s} ({\bf r} t;{\bf
r}^{\prime} t^{\prime}) + \int d{\bf r}_{1} \int d t_{1} \int d{\bf r}_{2} \int
d t_{2} \chi_{s} ({\bf r} t;{\bf r}_{1}t_{1}) K_{xc}({\bf r}_{1} t_{1}; {\bf
r}_{2} t_{2}) \chi ({\bf r}_{2} t_{2}, {\bf r}^{\prime} t^{\prime}) \nonumber
\end{equation}
$\chi_{s}$ is the magnetic response function of the Kohn-Sham
non-interacting system with the same unperturbed density $\rho_{0}$ as the full
interacting electron system, and $K_{xc}({\bf r} t;{\bf r}^{\prime}t^{\prime})
= (\delta b_{xc} ({\bf r},t)/\delta m ({\bf
r}^{\prime},t^{\prime}))|_{b=0,\rho_{0}}$
is the functional derivative of the effective
exchange-correlation magnetic field with respect to the induced magnetisation.
As emphasised in ref.\cite{Gross_review} eq.1 represents an exact
representation  of the linear magnetic response. The corresponding development for
systems at finite temperature in thermal equilibrium has also been described
\cite{Yang}. In practice
approximations to $K_{xc}$ must be made and this work employs the adiabatic
local approximation (ALDA) \cite{Gross_review} so
that $K^{ALDA}_{xc} ({\bf r} t;{\bf
r}^{\prime} t^{\prime}) = (d^{2}b_{xc}^{LDA}(\rho({\bf r},t),m({\bf r}, t))/
dm^{2}({\bf r},t))|_{\rho_{0},m=0} \delta ({\bf r} -{\bf r}^{\prime}) \delta
(t-t^{\prime}) = I_{xc}({\bf r}) \delta ({\bf r} -{\bf r}^{\prime}) \delta
(t-t^{\prime})$. On taking the Fourier transform with respect to time we obtain the
dynamic spin susceptibility $\chi ({\bf r}, {\bf r}^{\prime}; w)$.

For computational expediency we consider the corresponding {\it temperature}
susceptibility \cite{Fetter+Walecka} $\bar{\chi} ({\bf r}, {\bf r}^{\prime};
w_{n})$ which occurs in the Fourier representation of the temperature function
$\bar{\chi} ({\bf r} \tau;{\bf r}^{\prime} \tau^{\prime})$ that depends on
imaginary time variables $\tau$,$\tau^{\prime}$ and $w_{n}$ are the bosonic
Matsubara frequencies $w_{n}=
2 n \pi k_{B} T$. Now $\bar{\chi}( {\bf r}, {\bf
r}^{\prime}; w_{n}) \equiv \chi ({\bf r}, {\bf r}^{\prime}; i w_{n})$
and an
analytical continuation to the upper side of the real $w$ axis produces the
dynamic susceptibility $\chi ({\bf r}, {\bf r}^{\prime}; w)$. Using crystal
symmetry and carrying out a lattice Fourier transform we obtain the following Dyson
equation for the temperature susceptibility
\begin{equation}
\bar{\chi}({\bf x},{\bf x}^{\prime},{\bf q},w_{n})= \bar{\chi}_{s}({\bf
x},{\bf
x}^{\prime},{\bf q},w_{n}) + \int d {\bf x}_{1} \bar{\chi}_{s}({\bf x},{\bf
x}_{1},{\bf q},w_{n}) I_{xc}({\bf x}_{1}) \bar{\chi}({\bf x}_{1}, {\bf
x}^{\prime},{\bf q},w_{n})               \label{eq:temp}
\end{equation}
with ${\bf x}$,${\bf x}^{\prime}$ and ${\bf x}_{1}$ measured relative to
crystal lattice unit cells of volume $V_{WS}$.

In terms of the DFT Kohn-Sham
Green function of the static unperturbed
system
\begin{equation}
 \bar{\chi}_{s}({\bf x},{\bf x}^{\prime},{\bf q},w_{n})= -\frac{1}{\beta N}
Tr. \sum_{{\bf R}} \sum_{m} G({\bf x},{\bf x}^{\prime}, {\bf R}, \mu
+i\nu_{m}) G({\bf
x}^{\prime},{\bf x},-{\bf R},\mu + i(
\nu_{m}+ w_{n})) e^{i {\bf q} \cdot {\bf R}} \label{eq:G}
\end{equation}
where ${\bf R}$ is a lattice vector between the cells
from which ${\bf x}$ and ${\bf x}^{\prime}$ are
  measured, $\mu$ the chemical potential
and $\nu_{m}$ is a fermionic Matsubara
frequency $(2n+1) \pi k_{B} T$. The Green function can be obtained within the
framework of multiple scattering (KKR) theory \cite{Faulkner+Stocks}. This makes
this formalism applicable to
disordered alloys as well as ordered compounds and elemental metals, the
disorder being treated by the Coherent Potential Approximation (CPA)
\cite{KKRCPA}. Then the partially averaged Green function,
 $\langle G ({\bf r},{\bf r}^{\prime},z) \rangle_{{\bf r}\alpha, {\bf
r}^{\prime} \beta}$, where ${\bf r}$,${\bf r}^{\prime}$ lie within unit cells
occupied by $\alpha$ and $\beta$ atoms respectively, can be evaluated in terms
of deviations from the Green function of an electron propagating through a lattice
of identical potentials determined by the CPA ansatz \cite{Butler}.

To solve equation ~(\ref{eq:temp}), we use the direct method of
matrix inversion and local field effects are fully incorporated.
$\bar{\chi}({\bf q},{\bf q}; w_{n})=(1/V_{WS}) \int d {\bf x} \int
d {\bf x}^{\prime} e^{i{\bf q} \cdot({\bf
x} - {\bf x}^{\prime})} \bar{\chi}({\bf x},{\bf x}^{\prime}, {\bf q},
w_{n})$ can then be constructed.
The most computationally demanding parts of the calculation are the convolution
integrals over the Brillouin Zone which result from the
expression  for $\bar{\chi}_{s}$, eq.~(\ref{eq:G}).
Since all electronic structure quantities are
evaluated at complex energies, these convolution integrals have no
sharp structure and can be evaluated
straightforwardly by an application of adaptive quadrature \cite{Ben+Ezio}.

As discussed in ref. \cite{Fetter+Walecka}, for example, we can define the retarded
response function $\chi ({\bf q},{\bf q}, z)$ of a complex variable $z$. Since
it can be shown  \cite{Fetter+Walecka}
formally that $\lim_{z \rightarrow \infty} \chi (z) \sim 1/z^{2}$ and we can
obtain $\chi (i w_{n})$ from the above analysis it is possible to continue
analytically to values of $z$ just above the real axis, i.e. $z= w +i \eta$.
In order to achieve this we fit our data to a rational function
$\bar{\chi} ({\bf q}, {\bf q},w_{n}) = \chi ({\bf q})
(1+\sum_{k=1}^{M-2} U_{k}({\bf q}) w_{n}^{k})/(1+\sum_{k=1}^{M}
D_{k}({\bf q}) w_{n}^{k})$
with the choice of coefficients $U_{k}$,$D_{k}$ ensuring that the sum rule
involving the static susceptibility $\chi ({\bf q})$ is satisfied, i.e. $\chi
({\bf q})= (2/\pi) \int^{\infty}_{0} dw  Im \chi({\bf q},{\bf q},w)/
w$. We find that very good fits are obtained with small $M$.

For chromium and its alloys, we find that $M=2$ is sufficient to provide
excellent fits to the calculations of $\bar{\chi}$ over a wide range of
$w_{n}$'s, i.e. $\bar{\chi}^{-1} ({\bf q}, {\bf q},w_{n}) = \chi^{-1}
({\bf q})(1 + (w_{n}/\Gamma({\bf q})) + (w_{n}/\Omega({\bf q}))^{2})$ so that
$\chi^{-1}({\bf q}, {\bf q},w)= \chi^{-1} ({\bf q})(1 -i (w/\Gamma({\bf q}))
- (w/\Omega({\bf q}))^{2})$
(standard error $<$ 3$\%$ of mean).  For the
systems studied here  we find $\Omega({\bf q})/\Gamma({\bf q}) < 0.15 \ll 1$ and so
the {\it spin dynamics can be described in terms of a heavily
overdamped oscillator model}. Evidently $t_{SF}({\bf q}) = \hbar/\Gamma({\bf q})
$ represents a relaxation time for a damped diffusive spin fluctuation
with wavevector ${\bf q}$. Moreover,
the imaginary part of the dynamical susceptibility which, when multiplied by
$(1-\exp(-\beta w))^{-1}$, is
proportional to the scattering cross-sections measured in inelastic neutron
scattering experiments, is written
$Im \chi ({\bf q}, {\bf q},w) =\chi({\bf q}) w \Gamma^{-1}({\bf q})
/((1 -(w/\Omega({\bf q}))^{2})^{2} + (w/\Gamma({\bf q}))^{2})$. We note that
theories for the spin fluctuation effects upon itinerant
electron properties, including quantum critical phenomena, also invoke such a model
\cite{GGLetc}. The small ${\bf Q}$, $= ({\bf q} -{\bf q}_{0})$, dependence
of $\chi^{-1}({\bf q}_{0}+{\bf Q})$ and $\Gamma ({\bf q}_{0}+{\bf
Q})$ is of particular importance. (${\bf q}_{0}$ is where
$\chi^{-1}({\bf q})$ is smallest.)

Finite-temperature calculations were carried out for the static
susceptibilities of the three systems, using the experimental b.c.c. lattice
spacing of $Cr$, 2.88 $\dot{A}$, and von Barth-Hedin local exchange
and correlation \cite{vBH}.  We find that
(i) $Cr$ orders into an incommensurate AF state below 280K specified
by ${\bf q}_{0}=\{0,0,0.93\}$, where experiment yields $T_{N}= 311K$ and
${\bf q}_{0}=\{0,0,0.95\}$ \cite{Fawcett};
(ii) $Cr_{95}V_{5}$ does not develop magnetic order at any temperature,
as found in experiment \cite{Fawcett,Hayden}; and
(iii) $Cr_{95}Re_{5}$ orders into a weakly incommensurate AF state below
T=410K (${\bf q}_{0}=\{0,0,0.96\}$), whereas experimentally it forms a
commensurate AF state below $T_{N}$ of 570K \cite{Fawcett}.

Our calculations for $Im \chi ({\bf q}, {\bf q},w)$ are shown
in figures 1(a) and (b) for $Cr_{95}V_{5}$ and $Cr_{95}Re_{5}$ respectively. Our
calculation of
$Im \chi ({\bf q}, {\bf q},w)$ for paramagnetic $Cr$ at $T=300K$ is broadly similar
to that for paramagnetic $Cr$ at $T=0K$ by Savrasov \cite{SW+Sav} so a figure
is not presented. It shows incommensurate spin fluctuations
for small frequencies which are signified by
peaks in $Im \chi({\bf q},{\bf q},w)$ at
${\bf q}_{0}$ which is equal to the Fermi surface nesting vector
${\bf q}_{nest}$.
For increasing $w$ the peaks move to ${\bf q}=\{0,0,1\}$ i.e. the spin
fluctuations become commensurate. The spin fluctuations at 300K
shown in fig.1(a)
for $Cr_{95}V_{5}$, on the other hand, remain incommensurate up to much
higher frequencies maintaining intensity
comparable to that at the peak at low $w$. This qualitative difference between
the two systems has not been described before by a first-principles theory although
found experimentally \cite{Fawcett,Hayden}.
For lower temperatures we find that
$Im \chi({\bf q},{\bf q},w)$ of $Cr_{95}V_{5}$ becomes
a sharper function of $w$ and we can
also infer that $(1-\exp(-\beta w))^{-1} Im \chi({\bf q},{\bf q},w)$
should vanish for small $w$ when $T \rightarrow 0$K.
These aspects have also been noted from experimental
measurements \cite{Fawcett}.

It is striking that
the alloy's Fermi surface is well defined \cite{KKRCPA} despite
impurity scattering although it is more poorly nested
(the difference between the sizes of the
electron and hole octahedra is larger) than that of
$Cr$ owing to its fewer electrons. Once again the
  peaks in $Im \chi({\bf q},{\bf q},w)$ occur at the
  nesting vectors ${\bf q}_{nest} = \{0,0,0.9\}$. The
spin fluctuations in the paramagnetic phase of $Cr_{95}Re_{5}$ are shown in
fig.1(b). Here adding electrons by doping with $Re$ has improved the Fermi surface
nesting so that ${\bf q}_{nest} = \{0,0,0.96\}$ and $Im \chi ({\bf q},{\bf
q},w)$ has weight spread from
${\bf q}_{nest}$ to $\{0,0,1\}$. The dominant spin fluctuations now rapidly become
commensurate with increasing $w$. We obtain a rather similar picture from the
calculations for $Cr$ by artificially raising the chemical potential by a small
amount. Interestingly when
we account for thermally induced electron-phonon scattering
\cite{model} by adding a small shift ($\approx 20$ meV) to
the Matsubara frequencies $\nu_{m}$ in
eq.~(\ref{eq:G}), we find a tendency for the dominant spin fluctuations to
become commensurate at lower $w$ in both $Cr$ and $Cr_{95}Re_{5}$.

Some of these features also emerge qualitatively
from the simple parameterised models based on that part of the
band-structure near $\mu$ which leads to the nested electron and hole
octahedra at the Fermi surface \cite{model}.
Our `first-principles' calculations, being based on an
all-electron theory, however, need some additional interpretation. As
analysed by recent total energy calculations
\cite{Marcus}, b.c.c. $Cr$ with the experimentally measured lattice
spacing tends to form a commensurate AF phase at low
temperatures which is modulated by a spin density wave of appropriate
wavelength. The overall AF instability of the paramagnetic phase is promoted
by the approximate half-filling of the narrow 3d-bands \cite{Heine+Samson}
which is further modified by a weak perturbation
coming from the Fermi surface nesting. As dopants such as $V$ and $Re$
are added not only is the Fermi surface nesting altered but also the d-bands
become either further from or closer to being half-filled.

The calculations can be summarised in terms of the damped oscillator model.
$\chi^{-1} ({\bf q}_{0}+{\bf Q}) \simeq  \chi^{-1} ({\bf
q}_{0} + c Q^{2})$ for small
${\bf Q}$ for $Cr$ and $Cr_{95}V_{5}$ whereas for $Cr_{95}Re_{5}$,
$\chi^{-1} ({\bf q})$ is nearly constant for a range of ${\bf q}$ between ${\bf
q}_{nest}$ and $\{0,0,1\}$. We find the product $\gamma ({\bf q})$
of $\chi ({\bf q})$ with  damping factor $\Gamma({\bf q})$
to be only very weakly temperature
dependent in these three systems and $\simeq \gamma ({\bf q}_{0})$, a
constant, for small ${\bf Q}$,
yielding a dynamical critical exponent \cite{GGLetc}
of 2 typically assumed for antiferromagnetic itinerant electron
systems. The nature of the spin fluctuations can
be succinctly described via the variance $< m^{2}({\bf q})>$. From
the fluctuation dissipation theorem,
$< m^{2}({\bf q})> = (1/\pi) \int^{\infty}_{-\infty}
dw (1-\exp(-\beta w))^{-1} Im \chi ({\bf q},{\bf q}, w)$.
Fig.2 shows  $<
m^{2}({\bf q})>$ at several temperatures for $Cr$
where we have used a frequency cutoff
of 500 meV and so have not included the faster of the
quantum fluctuations.
Near $T_{N}$ the
magnetic fluctuations have their greatest weight around
the ${\bf q}_{nest}$.
At higher $T$ the peak diminishes and weight grows at
${\bf q}$'s nearer $\{ 0,0,1 \}$ reflecting the shift in the peak in
$Im \chi({\bf q},{\bf q},w)$ from ${\bf q}_{nest}$
to commensurate ${\bf q}$'s with increase in
frequency $w$. Similar plots to fig.2 for
$Cr_{95}V_{5}$ and $Cr_{95}Re_{5}$ show respectively a smaller and greater
tendency for the weight in $< m^{2}({\bf q})>$ to transfer in this way. If the
frequency cutoff is reduced, $< m^{2}({\bf q})>$ near $\{ 0,0,1 \}$ is
sharply diminished so that the Brillouin zone integral of $< m^{2}({\bf q})>$,
$< m^{2} >$,
decreases with increasing temperature as inferred from neutron scattering data
\cite{Fawcett}.

We have not included the effects of spin fluctuation interactions i.e.
mode-mode coupling \cite{MD+Moriya+LT} into our calculations and have
determined $T_{N}$ and the
static susceptibility by what is essentially an ab-initio Stoner theory. In weak
itinerant ferromagnets, for example, mode-mode coupling causes a dramatic
  suppression of the Curie temperatures from those estimated from a Stoner theory.
For the $Cr$ systems studied here, the
fair agreement with experiment which we obtain for $T_{N}$ and the relatively large
value of the damping factor $\Gamma ({\bf q})$ with respect to that in weakly
itinerant ferromagnets, is suggestive that such spin fluctuation effects are small.
Spin fluctuation calculations have, however, been carried out by Hasegawa
and others for simple
parameterised models of $Cr$
neglecting Stoner particle-hole excitations and Fermi surface nesting
\cite{Hasegawa}.
Using a functional
integral technique he made a high temperature (static) approximation so that
all the thermally induced fluctuations
were treated classically and found $T_{N}$ for
commensurate AF order to be 370K and $\sqrt{< m^{2} >}$ to increase linearly
with temperature above $T_{N}$.
A quantitative calculation, however, in which the Stoner particle-hole
excitations and spin fluctuations are treated within the same framework is
needed to determine unequivocally whether
or not a Stoner-like picture is adequate for these systems.

In summary, we have presented a first-principles framework for the calculation of
dynamic paramagnetic spin susceptibilities of metals and their alloys
at finite temperatures. At this
point we add that the approach can also be applied to the study of magnetic
excitations in magnetically ordered materials. The first applications
on the AF spin fluctuations in $Cr$ and $Cr_{95}Re_{5}$ above $T_{N}$ and in
paramagnetic $Cr_{95}V_{5}$ have found good agreement with available
experimental data.

We are grateful to F.J.Pinski, S.Hayden and R.Doubble for useful discussions.

\begin{figure}
\caption{$Im \chi({\bf q},{\bf q},w)$
of (a)  $Cr_{95}V_{5}$ at
$T=$300K and (b) $Cr_{95}Re_{5}$ at $T=1.09 T_{N}$ (450K) in units of $\mu_{B}^{2}
eV^{-1}$ for wave-vectors ${\bf q}$ along the $\{ 0,0,1 \}$ direction where
${\bf q}$ is in units of $\pi/a$ (a is
the lattice spacing). The frequency axis, $w$, is marked in meV.}
\end{figure}
\begin{figure}
\caption{ The variance of the spin fluctuations in $Cr$ $<
m^{2}({\bf q})>$ in $\mu_{B}^{2}$ for
wave-vectors ${\bf q}$ along the $\{ 0,0,1 \}$ direction at
350K (full line),400K (long dashes),450K (medium dashes)
and 700K (short dashes).}
\end{figure}
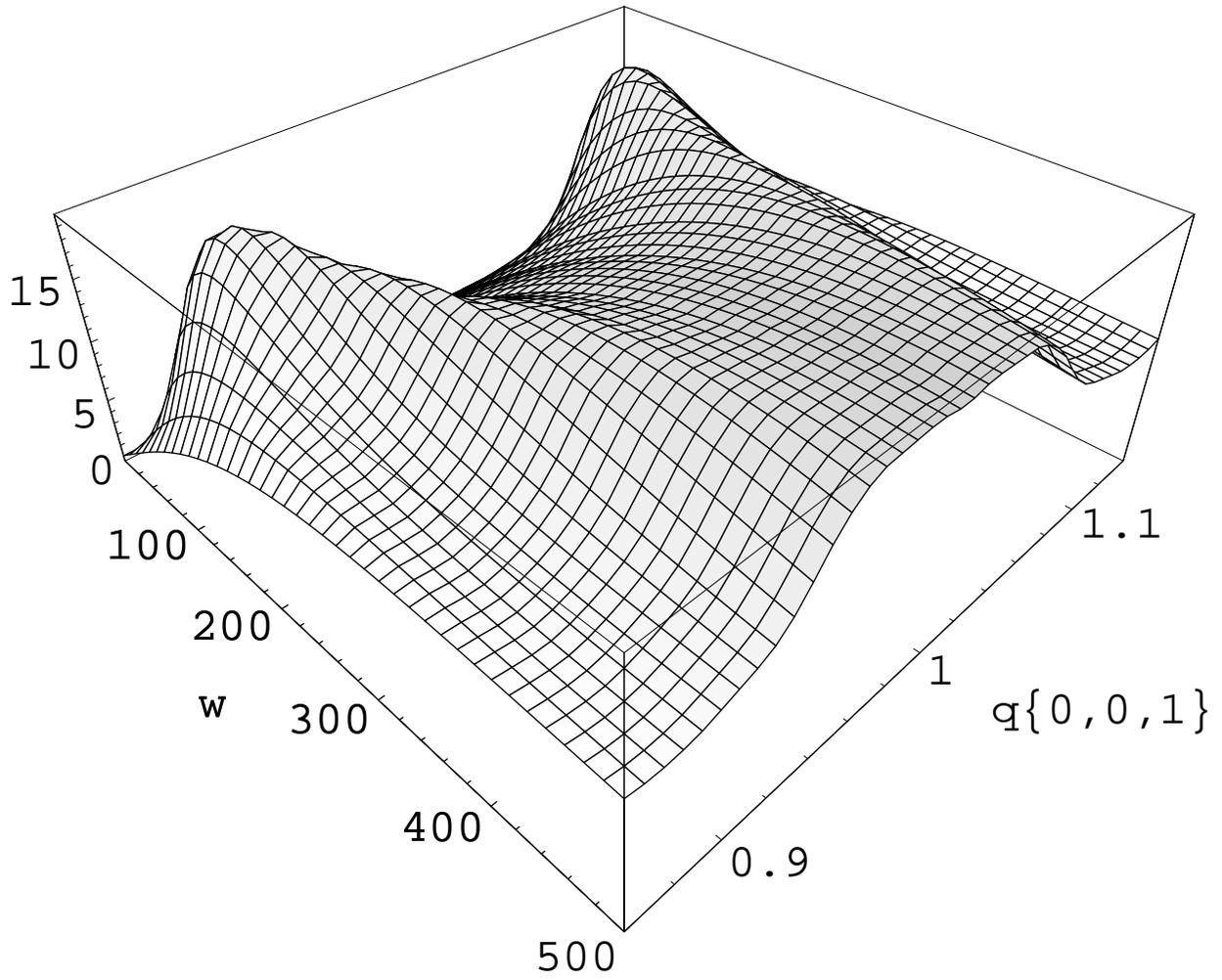
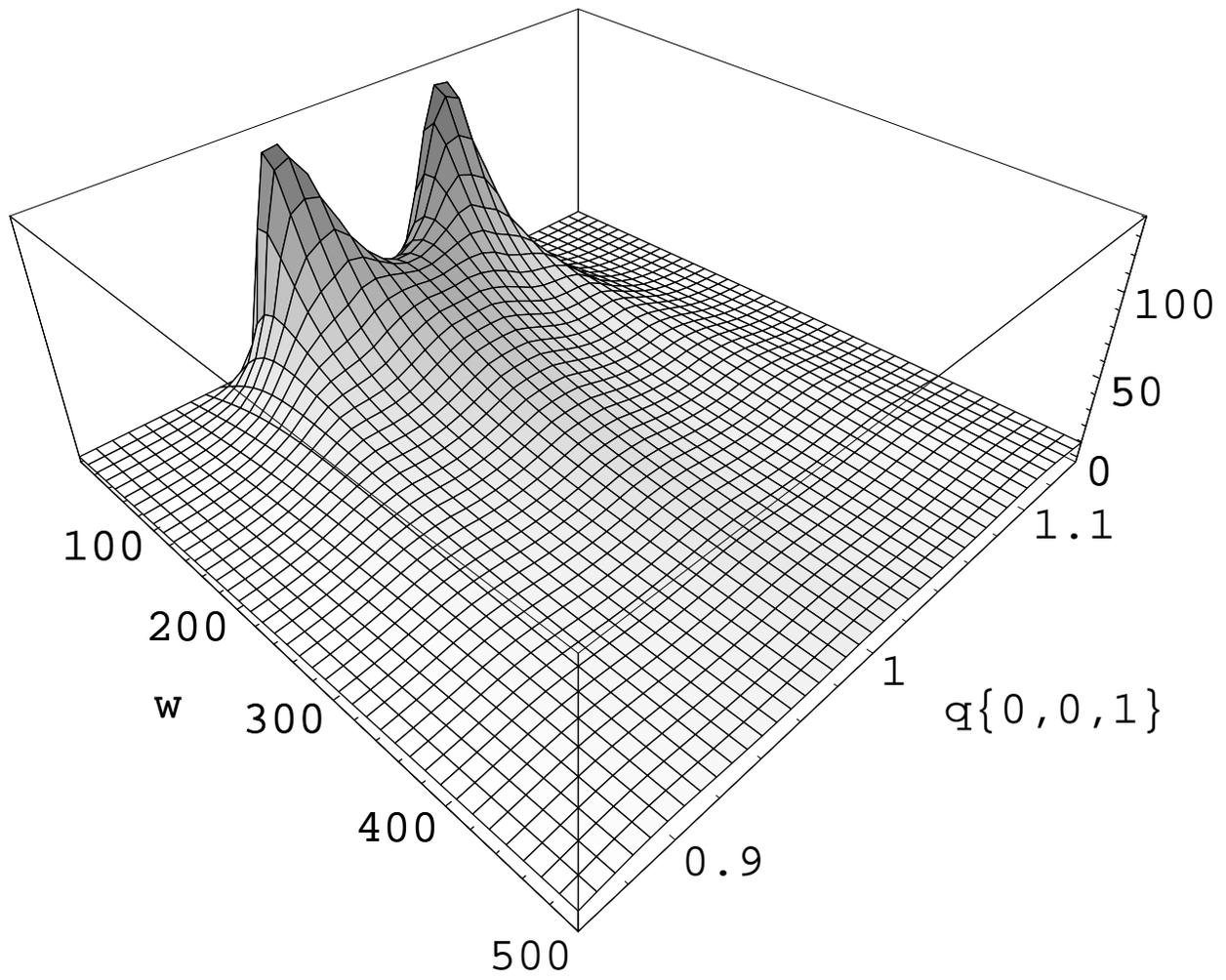
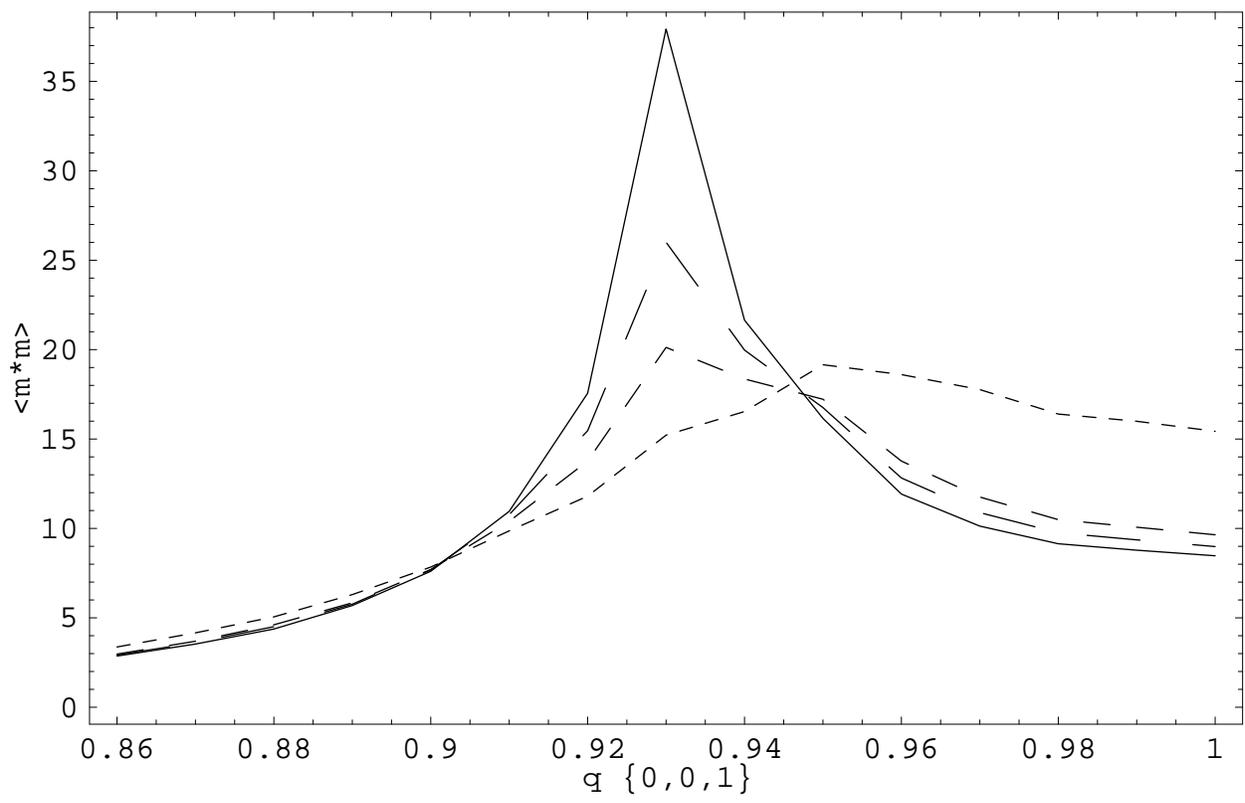
\end{document}